\documentclass[twocolumn]{aa}
\usepackage{graphicx}
\usepackage{epsfig}
\begin{document}
%\headnote{Research Note}

\title{The 2003 eclipse of EE Cep is coming}

\subtitle{A review of past eclipses}

\author{ D. Graczyk$^{1}\!\!$, M. Miko{\l}ajewski$^{1}\!\!$, T. Tomov$^{1}\!\!$, D. Kolev$^{2}$ \and I. Iliev$^{2}$}

\offprints{M. Miko{\l}ajewski}
\institute{$^{1}$ Centre for Astronomy, ul.~Gagarina 11, 87-100 Toru\'n, Poland \\
           $^{2}$ National Astronomical Observatory Rozeh, Institute of Astronomy, BAS, PO Box 136, 4700 Smolyan, Bulgaria\\
\email{weganin@astri.uni.torun.pl}
\email{mamiko@astri.uni.torun.pl}}

\date{} 
\abstract{EE Cep is an eclipsing binary with a period of 5.6 years.
The next eclipse will occur soon, in May--June 2003, and all available past
eclipses were collected and briefly analysed. EE Cep shows very large
changes of the shape and the depth of minima during different eclipses,
however it is possible to single out some persistent features. The analysis
suggests that the eclipsing body should be a long object surrounded by an
extended semi-transparent envelope. As an explanation, a model of a
precessing optically thick disc, inclined to the plane 
of the binary orbit, is invoked. The changes of
its spatial orientation, which is defined by the inclination of the disc and
the tilt, induced most probably by precession of the disc spin axis 
with a period of about 50 years, produce strange photometric 
behaviour of this star. The H$\alpha$
emission, and possibly the NaI absorptions, show significant changes during
several months outside of the eclipse phase.
}
\maketitle

\markboth{D. Graczyk et al.: The 2003 eclipse of EE Cep is coming}{D.
Graczyk et al.: The 2003 eclipse of EE Cep is coming}

\section{Introduction}

EE Cep (BD+55\degr 2693) is an extraordinary long-period eclipsing binary
showing very large and complex changes of the shape and the depth of the
minima (e.g.~Baldinelli \& Ghedini~1977). Quite recently Miko{\l}ajewski \&
Graczyk (\cite{MG} - hereafter MG) have suggested that the reason for these
changes is the presence of a non-stellar eclipsing body in the system -- a
precessing, opaque and dark disc.

This star was identified as a variable star by Romano (\cite{roman56}) in
1952 and it was confirmed by Weber (\cite{weber56}) who reported
observations of the previous minimum in 1947. The colour changes during the
1952 minimum were very small and the star was originally classified by
Romano as an R CrB variable. Perova (\cite{pero57}) reported about 500
photographic observations of EE Cep between 1899-1908 and 1933-1955 but no
brightness changes were recorded. Romano \& Perissinotto (\cite{roman66})
have suggested that it may be an eclipsing binary but they did not give any
approximate ephemeris. Its binary nature was confirmed later by Meinunger
(\cite{mein73}) who analysed over 500 photographic plates from Sonneberg
Observatory. The period of the star is almost 2050 days and only the primary
minimum can be detected. Meinunger has not detected any changes of the
brightness between the 1958 and 1964 minima. The brightness of the star
outside of the eclipse phases is constant within the observational errors:
$V_{\rm max}=10.78$ (Meinunger \& Pfau \cite{mein81}, MG).  The primary
component is apparently a normal B5 III star (Herbig \cite{her60},
Baldinelli et al.~\cite{bal81}). The emission visible in $H_{\rm \alpha}$ and
$H_{\rm \beta}$ lines arises most likely from an invisible companion (Br{\"u}ckner
\cite{bryk76}). The absorption spectrum of the companion is never seen.

The early model of EE Cep presented by Meinunger (1976) assumed that the
companion should be a cool star of M type, much less luminous but larger
than the primary B5 III star, surrounded by a strongly extended envelope.
The pulsations of the cool star would change its diameter and thus lead to
the changes of the minima duration and depth. The envelope would cause an
atmospheric eclipse in the beginning and in the end of the minimum,
producing typical wings in the light curves. However, such a simple model
cannot explain some of the features observed in EE Cep, as was argued by MG
and in this paper.

\begin{figure*}
\begin{minipage}{0.49\linewidth}
\resizebox{\linewidth}{!}{\includegraphics{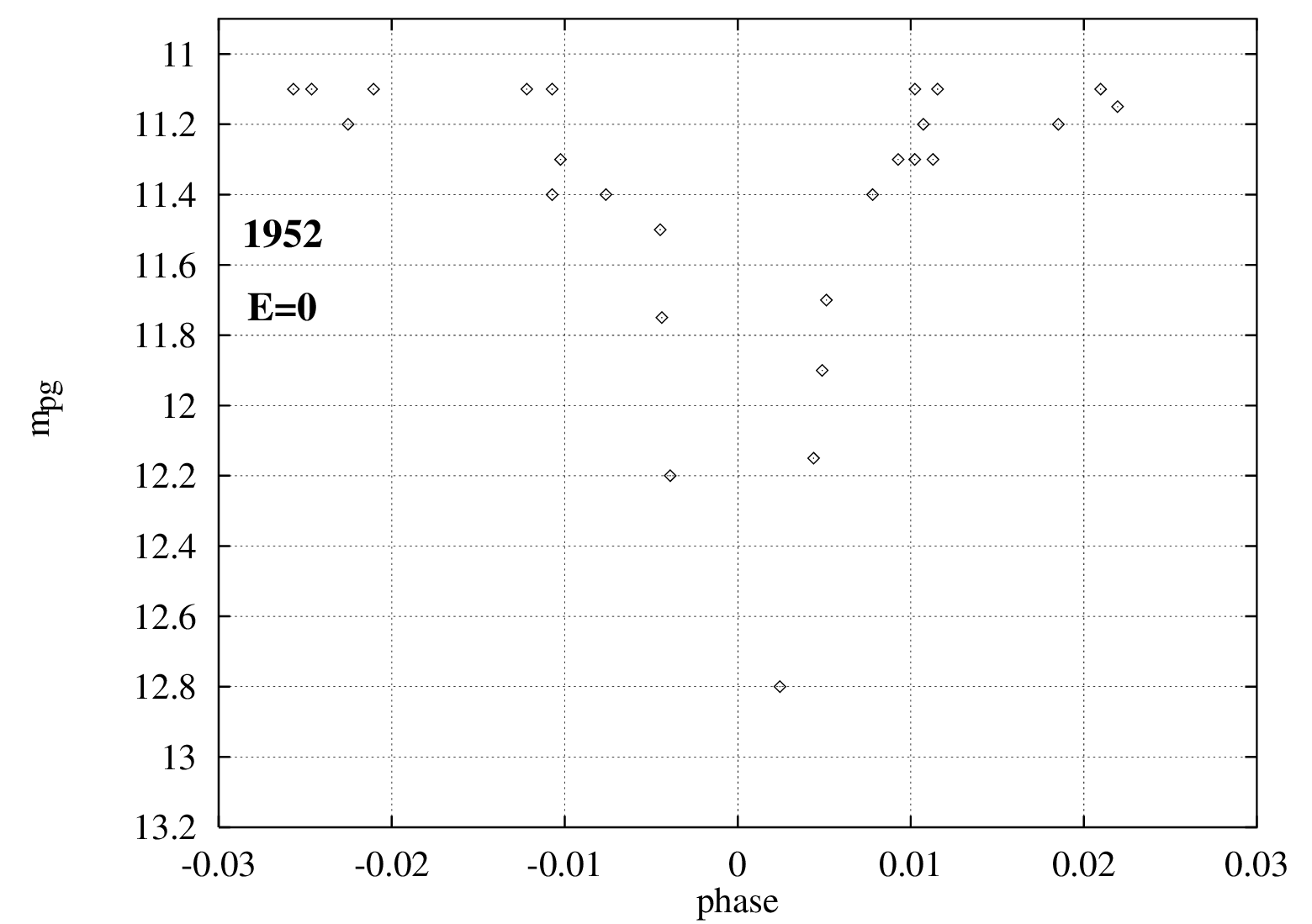}}
\mbox{}\\[-0.2cm]
\resizebox{\linewidth}{!}{\includegraphics{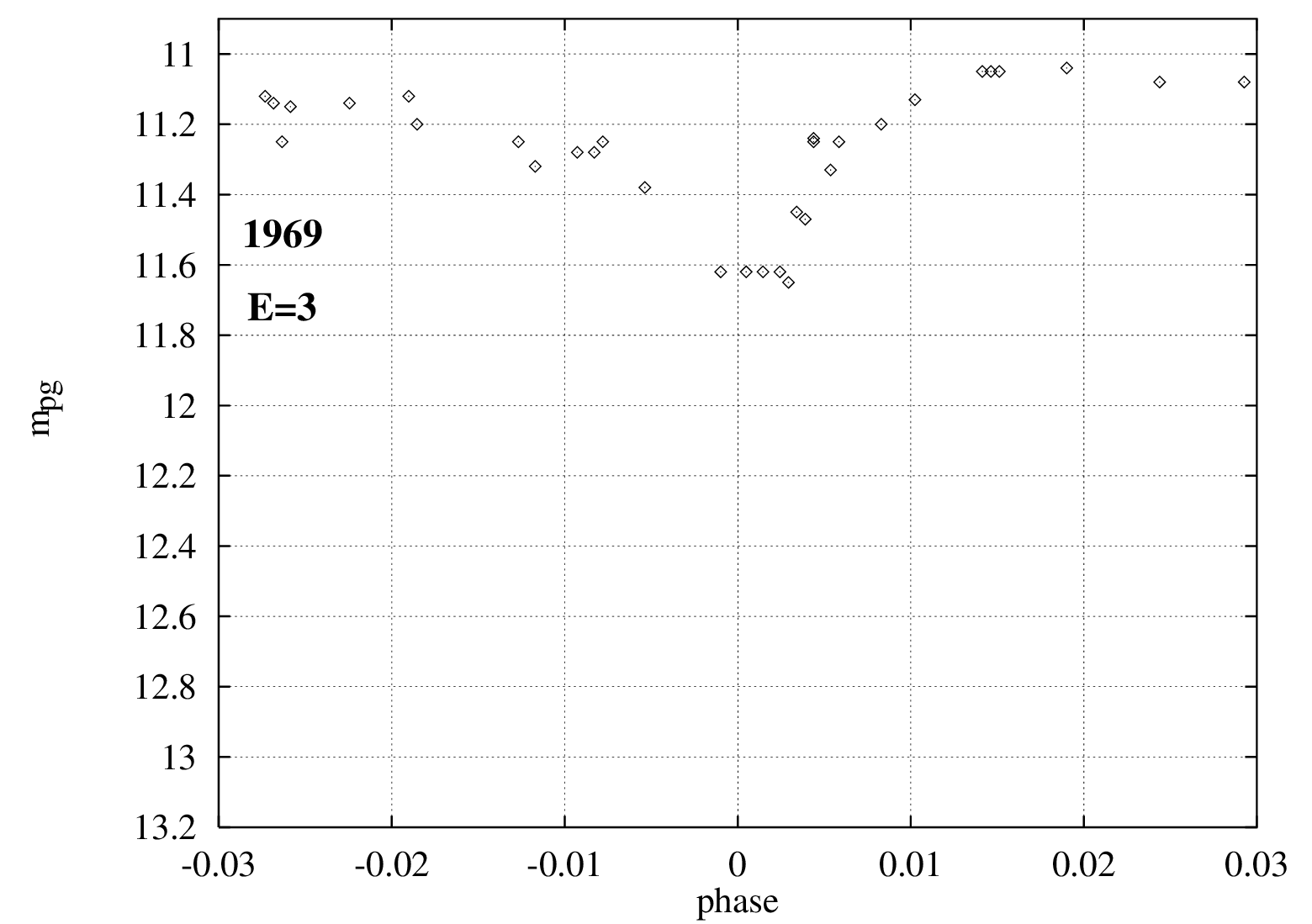}}
\mbox{}\\[-0.2cm]
\resizebox{\linewidth}{!}{\includegraphics{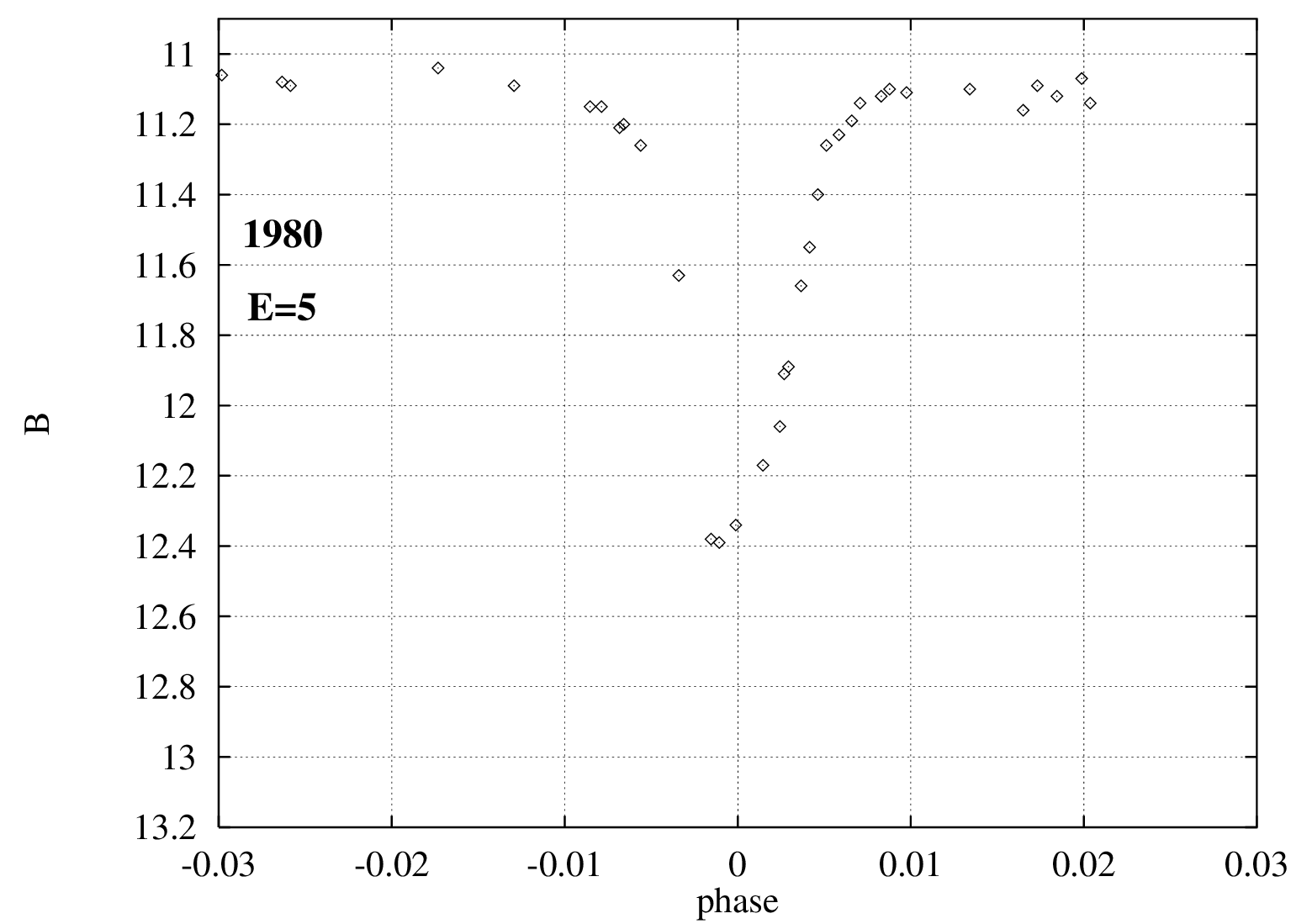}}
\end{minipage}\hfill
\begin{minipage}{0.49\linewidth}
\resizebox{\linewidth}{!}{\includegraphics{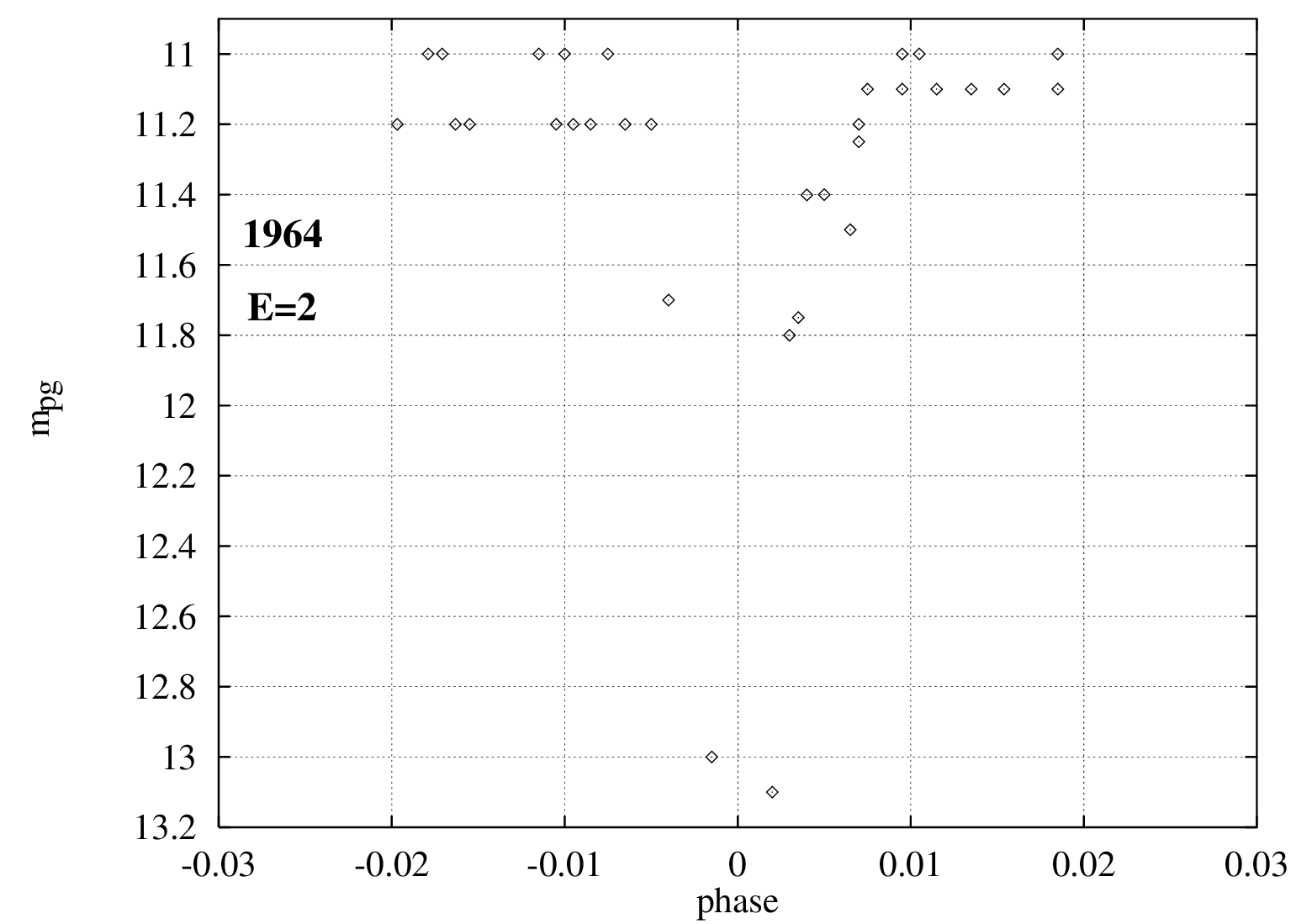}}
\mbox{}\\[-0.2cm]
\resizebox{\linewidth}{!}{\includegraphics{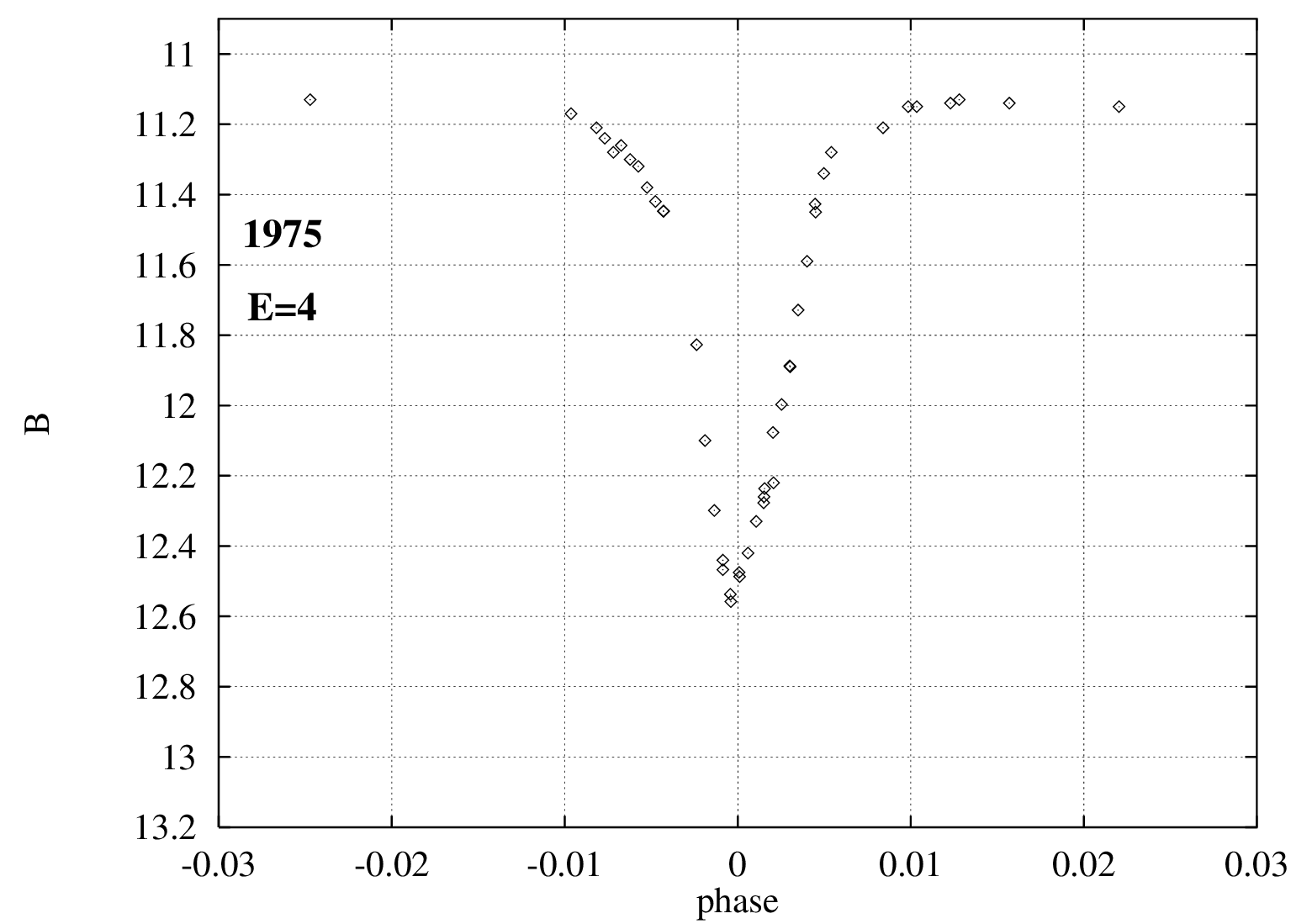}}
\mbox{}\\[-0.2cm]
\resizebox{\linewidth}{!}{\includegraphics{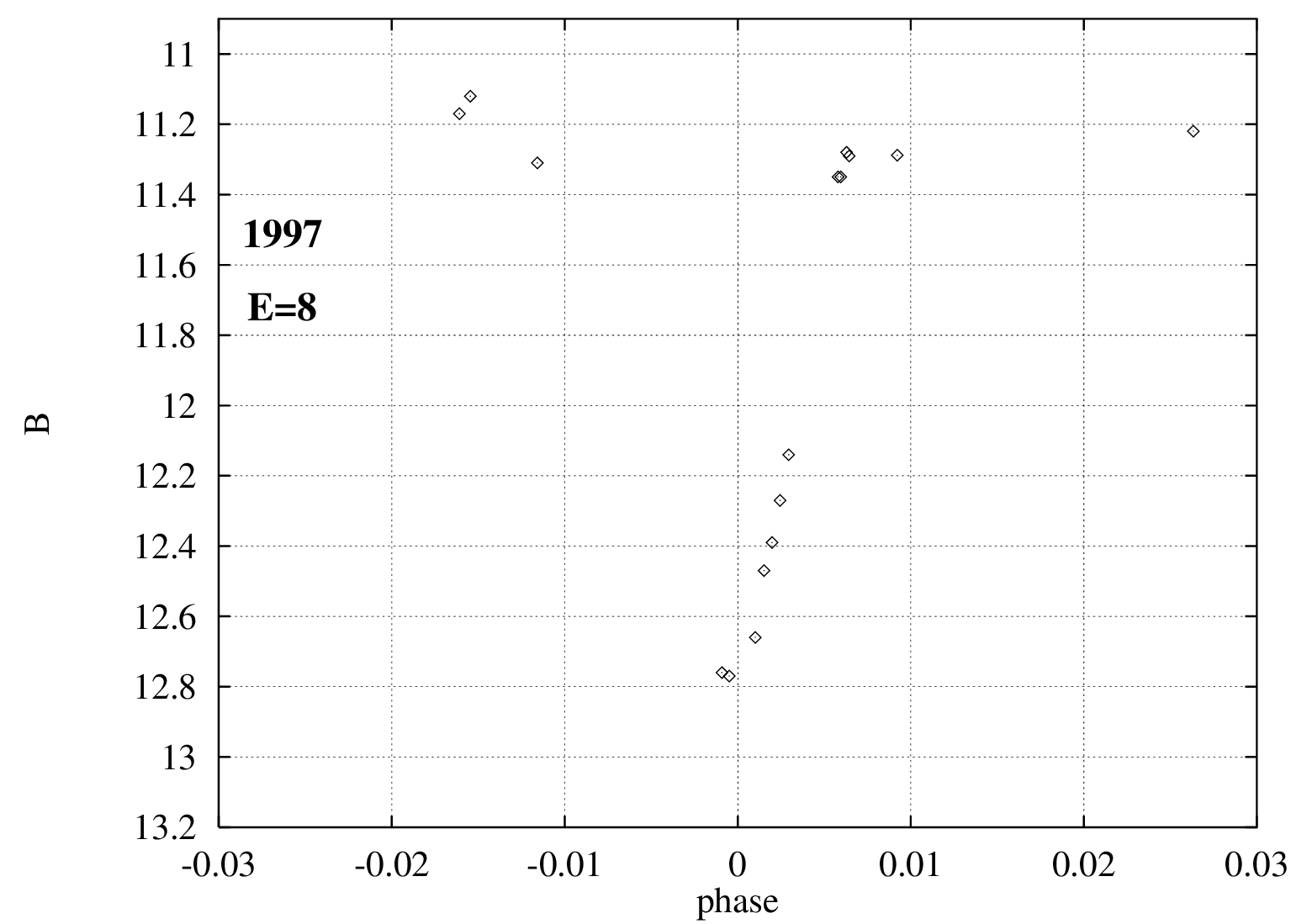}}
\end{minipage}
\caption{EE Cep's eclipses in $B/m_{\rm pg}$ light.}
\label{fig:obrazb}
\end{figure*}

\begin{figure*}
\begin{minipage}{0.49\linewidth}
\resizebox{\linewidth}{!}{\includegraphics{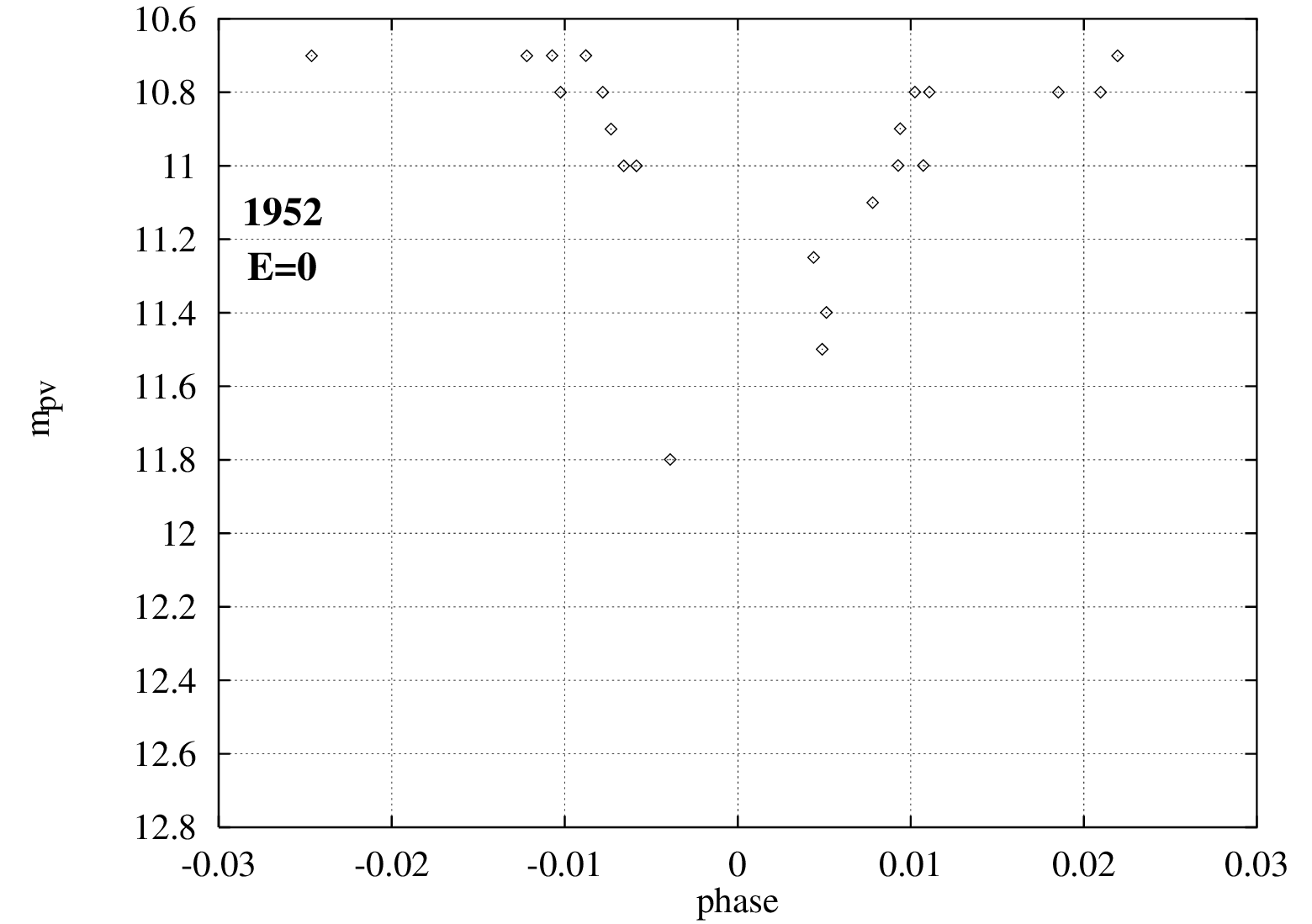}}
\mbox{}\\[-0.2cm]
\resizebox{\linewidth}{!}{\includegraphics{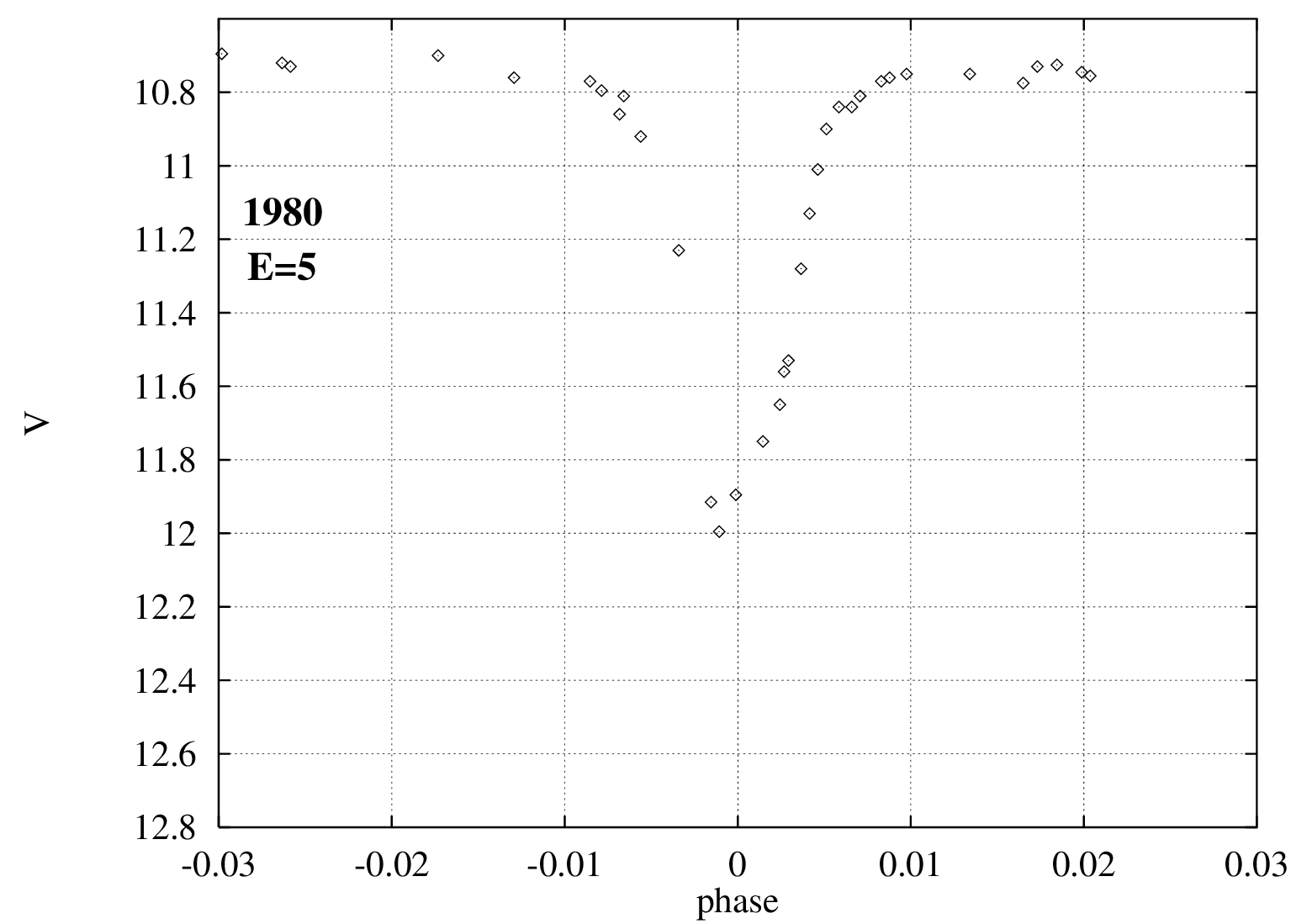}}
\mbox{}\\[-0.2cm]
\hspace*{0.5cm}\resizebox{0.93\linewidth}{!}{\includegraphics{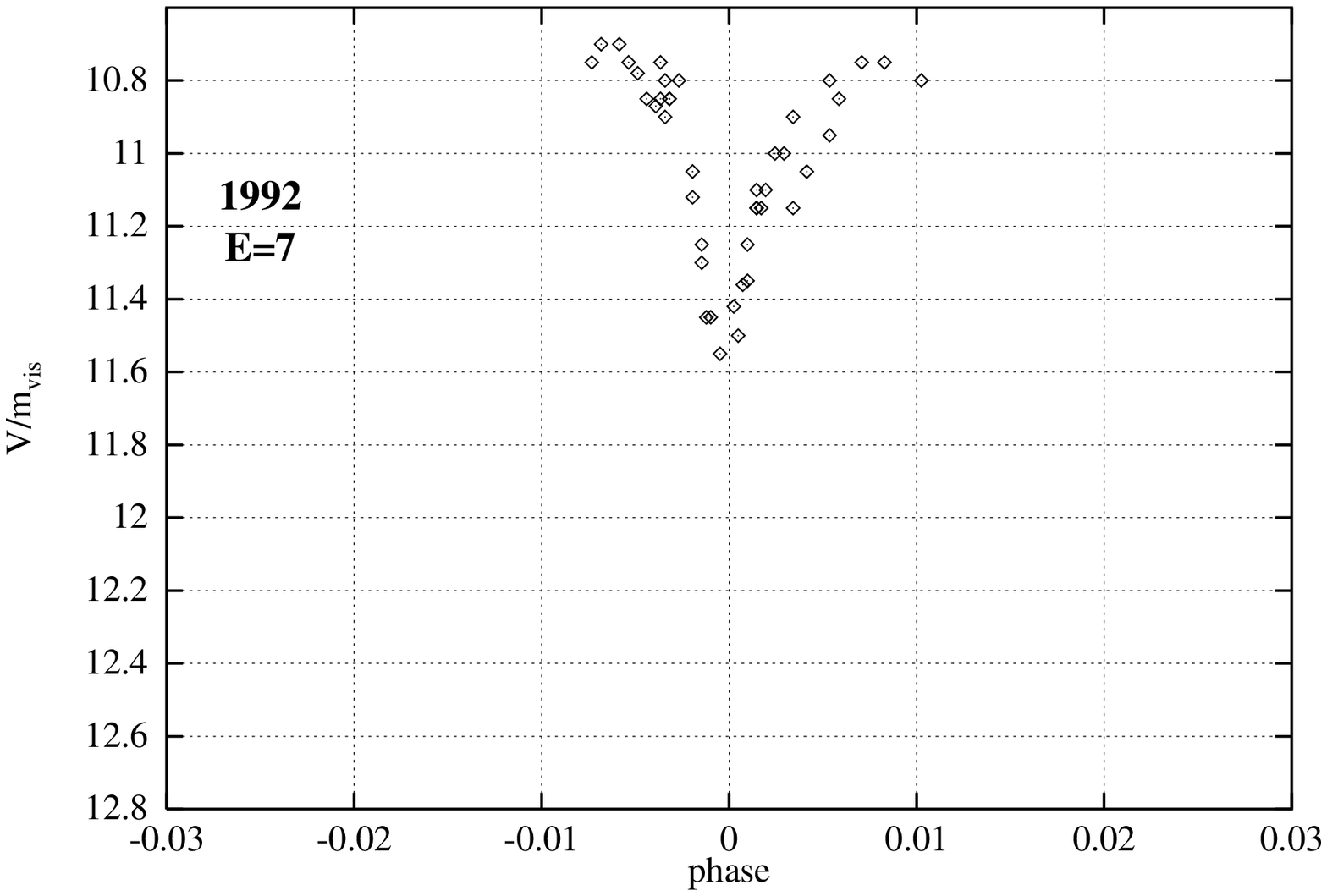}}
\end{minipage}\hfill
\begin{minipage}{0.49\linewidth}
\resizebox{\linewidth}{!}{\includegraphics{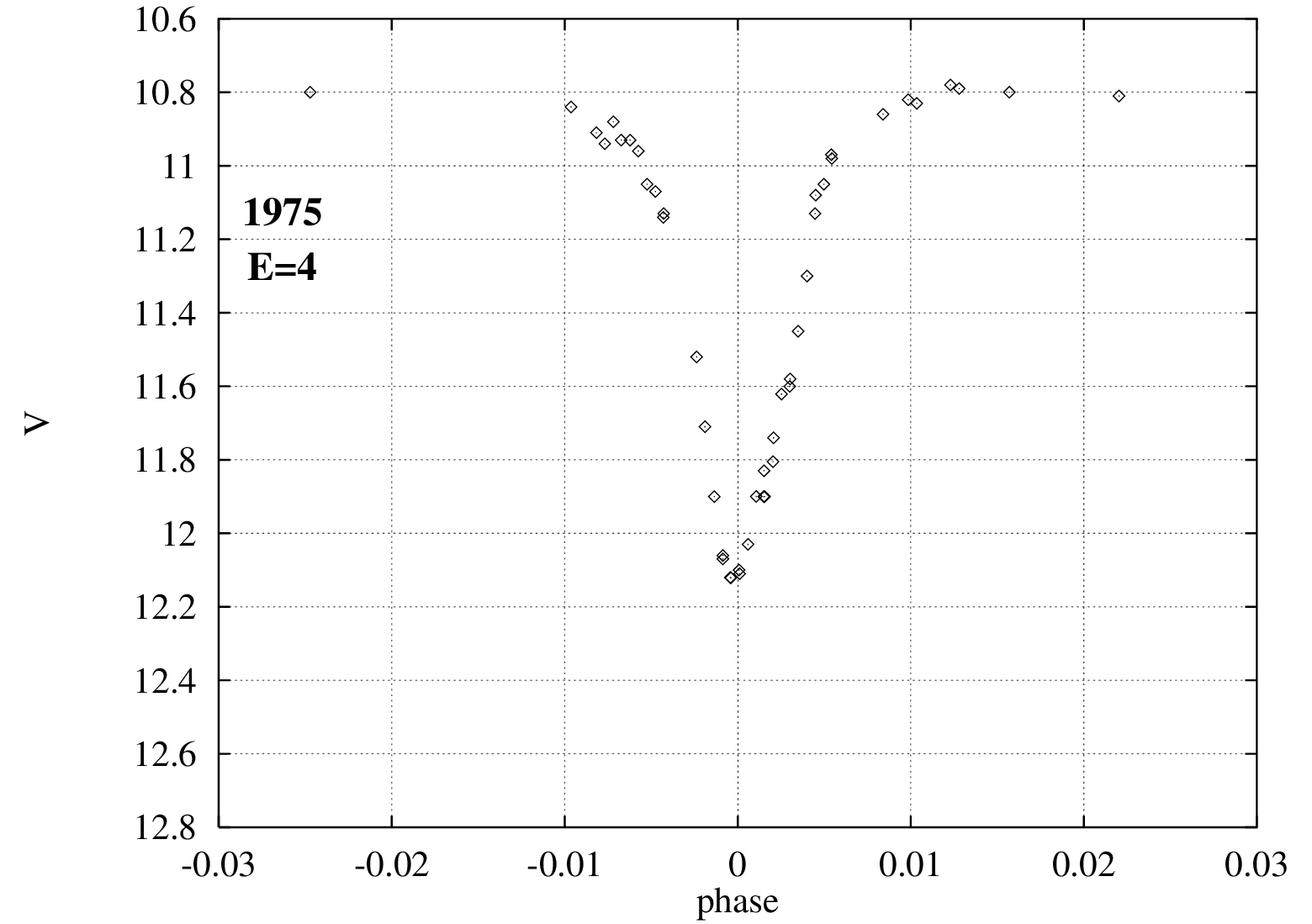}}
\mbox{}\\[-0.2cm]
\resizebox{\linewidth}{!}{\includegraphics{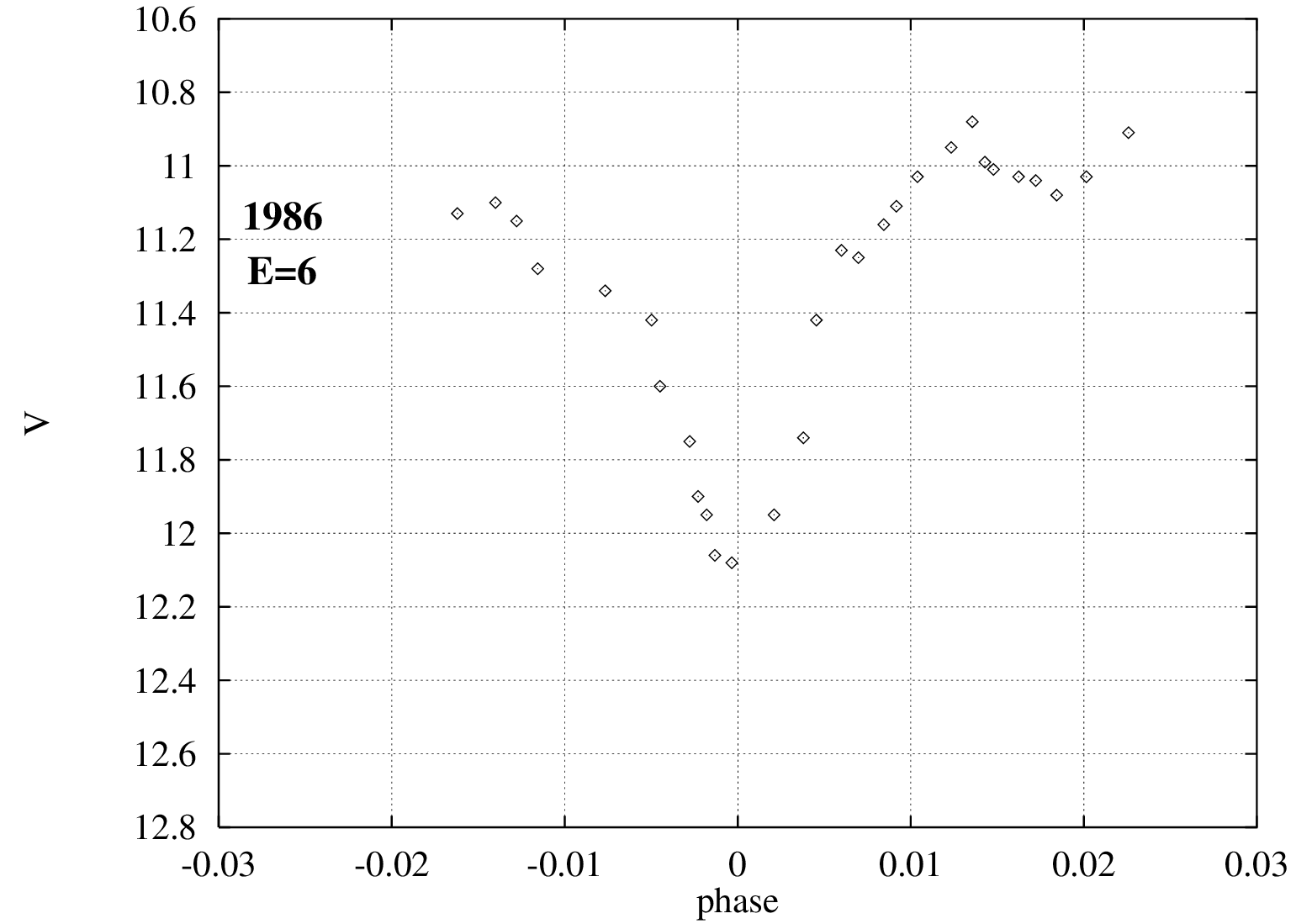}}
\mbox{}\\[-0.2cm]
\hspace*{0.5cm}\resizebox{0.93\linewidth}{!}{\includegraphics{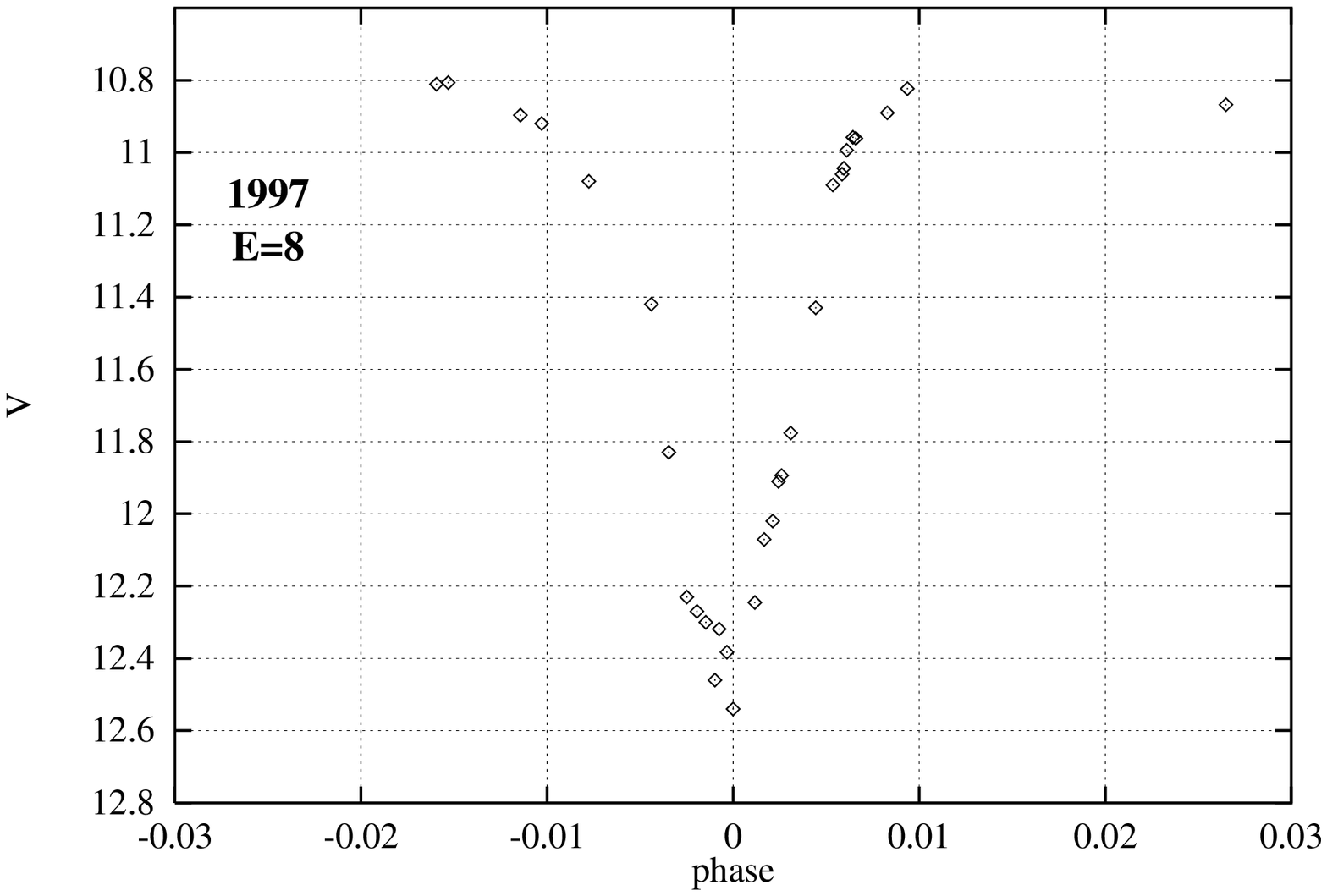}}
\end{minipage}
\caption{EE Cep's eclipses in $V/m_{\rm pv}/m_{\rm vis}$ light.}
\label{fig:obrazv}
\end{figure*}

\section{Photometric behaviour}

\subsection{A review of past eclipses}        

The typical minimum of EE Cep has large wings, the duration of the eclipse
is $D\approx 40$ days or a phase interval~$\Delta P=0.02$, and the depth of the eclipse is
moderate $\Delta B \approx 1.5$ mag. Such minima were observed in 1975, 1980
and 1997. But some eclipses observed up to date deviated strongly from this
picture. We have collected observations of all previously observed eclipses of EE Cep
in two roughly similar photometric wavelength bands: 1) $B$ and $m_{\rm pg}$;
2) $V$, $m_{\rm pv}$ and $m_{\rm vis}$. The references used for the
photometry are listed in Table\ref{tab:korel}, whereas the data are
presented in Figs.~\ref{fig:obrazb} and~\ref{fig:obrazv}. Part of the
historical data were published only as diagrams and they needed to be read
from these figures. All observations were phased using the ephemeris given
by MG.

\begin{table*}
\begin{minipage}[t]{\linewidth}
\caption{The duration and the depth of eclipses.}
\tabcolsep5pt
\label{tab:korel}
\begin{tabular}{*{5}{c}l}
\hline\hline
Epoch & Year & $\Delta$P & $\Delta V/m_{pv}/m_{vis}$ & $\Delta B/m_{pg}$ & References (source of data)\\ \hline
$\!\!\!-1$&1947&--& --   &1.5: & Weber \cite{weber56} (table)\\
0 & 1952 & 0.020$\pm$0.002 &$>1.2$&1.7$\pm$0.2 & Romano \cite{roman56} (figure)\\ 
1 & 1958 & 0.015$\pm$0.003 &--  &1.9$\pm$0.2 & Romano \& Perissinotto \cite{roman66} (table)\\
2 & 1964 & 0.015$\pm$0.002 &--  &2.1$\pm$0.2 & Meinunger \cite{mein73} (figure)\\ 
3 & 1969 & 0.030$\pm$0.003 &--  &0.6$\pm$0.1 & Baldinelli et al.~\cite{bal75} (table)\\ 
4 & 1975 & 0.021$\pm$0.001 &1.34$\pm$0.02&1.43$\pm$0.02 & Meinunger \cite{mein76} (table), Zaitseva et al. \cite{zait75} (table)\\ 
5 & 1980 & 0.019$\pm$0.001 &1.25$\pm$0.02&1.30$\pm$0.02 & Baldinelli et al. \cite{bal81} (figure), Meinunger \& Pfau \cite{mein81} (table)\\ 
6 & 1986 & 0.025$\pm$0.002 &1.1$\pm$0.1& --  & Di Luca \cite{luca88} (figure)\\ 
7 & 1992 & 0.011$\pm$0.002 &0.85$\pm$0.15& --  & Halbach \cite{hal92} (figure)\\
8 & 1997 & 0.023$\pm$0.001 &1.61$\pm$0.02&1.68$\pm$0.02 & MG, Halbach \cite{hal99} (figure) \\ \hline
\end{tabular}
\end{minipage}
\mbox{}\\
\begin{minipage}{\linewidth}
Note: the total duration was calculated as time period between $1\!^-\!\!$st
and $4\!^+\!$th external moments (see \S 2.2) and have been expressed as a
part of the orbital period ($\Delta P$). Errors were estimated according to
the accuracy of the data and the quality of the light curve covering. 
\end{minipage} 
\end{table*}

The 1947 minimum ($E=-1$) was the first one detected and its depth was 1\fm
5 in $m_{\rm pg}$. There is no information about the duration and shape of the
eclipse and it is not shown in Fig.~\ref{fig:obrazb}. The 1952 minimum
($E=0$) was deep (at least 1.7 mag in $m_{\rm pg}$) and has an asymmetric shape.
During the next eclipse, in 1958 ($E=1$), there are only two observational
points and two upper limits $\approx 13$\fm in $m_{\rm pg}$ at the bottom of the
minimum (Romano \& Perissinotto 1966), therefore we also omitted this
minimum in Fig.~\ref{fig:obrazb}. Nevertheless, it could be very similar, or
maybe a little deeper, to the following one in 1964 ($E=2$). The depth of
these two minima is at least 2 mag. The duration of the 1964 eclipse was
only 30 days.

During the 1969 ($E=3$) eclipse, the minimum was, in contrast to the
previous one ($E=2$), extremely shallow ($\Delta m_{\rm pg} = 0.6$ mag) and a
flat bottom phase was clearly observed. Also, the 1969 eclipse was probably
the longest one -- its total duration was about 60 days. The 1975 ($E=4$)
and 1980 ($E=5$) eclipses were more or less typical, while the 1986 ($E=6$)
minimum was again shallow and strongly asymmetric. The 1992 ($E=7$) eclipse
was observed visually by three amateur astronomers from AAVSO and seven $VR$
measurements with a CCD camera have been obtained by J.~Borovicka from
Ondrejov Observatory in the Czech Republic (Halbach \cite{hal92}).
Unfortunately, there is a zero-point offset of +0\fm 16 between Borovicka's
$V$ magnitude scale and Meinungers's (\cite{mein76}) and MG's scales. Also,
the visual light curves strongly suffer from differences in zero-point
scales. Thus we have shifted visual estimations to the corrected CCD
observations, although three points still lie out of the light curve
(Fig.~\ref{fig:obrazv}). This minimum was shallow ($\Delta m_{vis} \approx
0.8$) and probably the shortest one -- its total duration was about 22 days.
The last observed 1997 ($E=8$) eclipse was, in contrary, deeper than usual.
However, its shape and the total duration were similar to those of 1975 and
1980.

\subsection{The characteristic phases of the minimum}

In general we can note that 1) the shape of the eclipse may change from the
deep and narrow minimum (like in 1964) to a shallow and very wide minimum
(1969, 1986), 2) a flat-bottom phase may occur which is characteristic for
annular eclipses, 3) wide wings are often visible and 4) there is a
persistent asymmetry of the light curve between the descending and the
ascending branches of the minima.

\begin{figure}
\begin{minipage}{\linewidth}
\includegraphics[width=\textwidth]{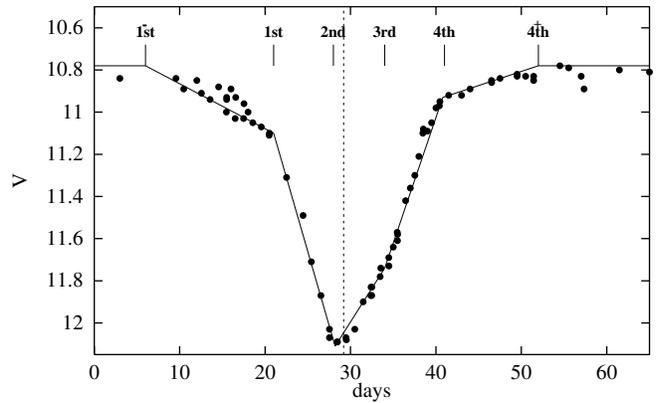}
\caption{The characteristic shape of the EE Cep's minima (E=4 and E=5 joined, se for details MG). In general six moments can be distinguished in most of minima. The wings corresponding to semi-atmospheric eclipse are apparent between $1\!^-\!\!$st-1st and 4-$4\!^+\!$th moments. The ``true'' eclipse can be defined as the part of the light curve between the 1st and 4th moments, with a sloping-bottom between 2nd and 3rd contacts.} 
\label{fig:term.ps}%
\end{minipage}
\end{figure}

Although the light curve of EE Cep shows these changes, it is possible to
single out the most typical shape of the minimum. MG suggested that two
eclipses from 1975 and 1980 are most typical. Basing on the data from these
two eclipses they construct the representative ("template") eclipse shown in
Fig.~\ref{fig:term.ps}. There are some typical features of the minimum which
are also observed in other eclipses of EE Cep. The most important is the
presence of six characteristic moments during an eclipse. Moments 1st, 2nd,
3rd and 4th correspond to contacts observed in normal eclipsing binaries,
while moments $1\!^-\!\!$st and $4\!^+\!$th denote the start and the end of
the phase of semi-atmospheric eclipse, respectively. Another feature is the
asymmetry of the light curve between the ascending and descending branches
of the minimum. Such asymmetry is clearly visible in eclipses of epochs 4, 5
and 8 and may be supposed to occur in the eclipse of epoch 6. The asymmetry
seems to be caused by the presence of the sloping-bottom phase (between 2nd
and 3rd contacts -- Fig.~\ref{fig:term.ps}) which influences the shape of
the ascending branch. This picture is supported by the fact that during
eclipse of epoch 3 four distinct contacts were observed and the bottom was
flat. Thus the part of the light curve between the 2nd and 3rd moments is
not part of the egress.

The characteristic shape and phases of the eclipse are independent of the
photometric band. Over a broad spectral range, from $U$ until $I$, the
eclipse seems to be almost grey (MG). During three eclipses (E = 4, 5, 6)
the depth in the $B$ band was on average $0.07$ mag deeper than in the $V$
band, possibly due to a selective extinction of the eclipsing body matter.
 
\subsection{Depth -- duration relation}

\begin{figure}
\centering
\includegraphics[width=0.5\textwidth]{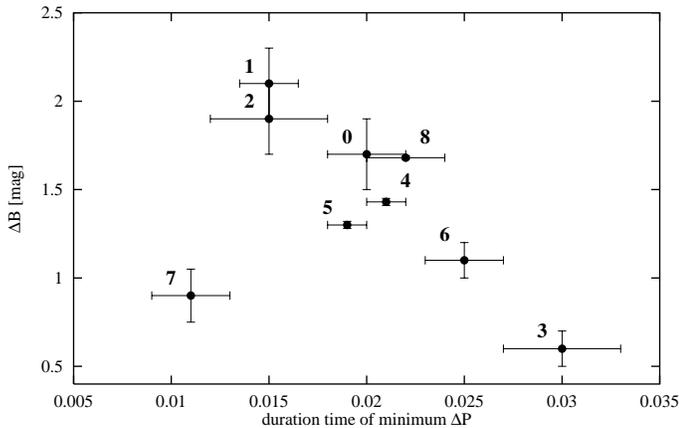}
\caption{The depth--duration relation for EE Cep's eclipses. Each eclipse is signed by
its epoch number.}
\label{fig:korel}
\end{figure}

Table~\ref{tab:korel} presents the total duration and the depth of EE Cep's
eclipses estimated from data in Figs.~\ref{fig:obrazb} and~\ref{fig:obrazv}.
Figure~\ref{fig:korel} presents the relation between the duration and the
depth of eclipses in $B/m_{\rm pg}$. For epochs 6 and 7 we have only $V/vis$
estimates so we applied a correction of $\Delta m=+0.07$ mag to the depth of
these two eclipses. It is worth noting that there is a quite distinct
correlation between these two parameters -- longer eclipses correspond to
shallower minima. Also, from the inspection of
Figs~\ref{fig:obrazb}~and~\ref{fig:obrazv} one can note that the shallowest
minima have the most asymmetric wings. The minimum of Epoch 7 seems to be an
exception to the rule. However, the estimates of $\Delta P$ and $\Delta V$
for this eclipse are based mainly on inconsistent visual observations, this
may reflect the presence of a more complex relation than simple linear
dependence.

\begin{figure}
\centering
\includegraphics[width=0.5\textwidth]{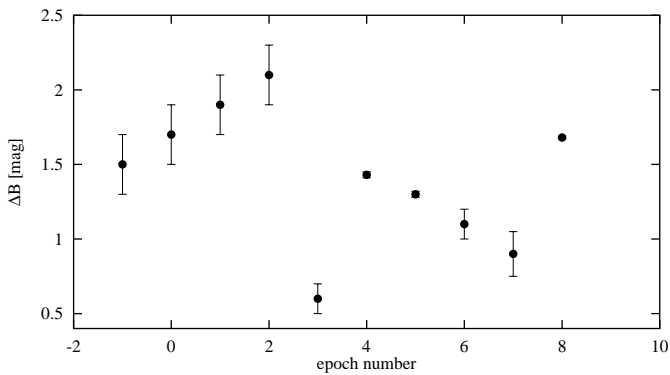}
\caption{Time dependence relation for the depth of the eclipses.}
\label{fig:order}
\end{figure}

The value of $\Delta B/m_{\rm pg}$ depths versus the number of epoch is
presented in Fig.~\ref{fig:order}. Its evolution is rather complex, but a
possible period of nine epochs ($\sim 50$ years) can be suggested. Because
it is almost equal to the timespan of all the historical data, new
observations are needed to confirm its existence. The duration of the
eclipses, which is a more difficult parameter to measure, do not show such
clear time evolution.
 
\section{The model of disc eclipses}

The model presented by MG assumes that the invisible component is an
optically thick, dark disc inclined to the orbital plane -- a ``cousin'' of the disc in
$\epsilon$ Aur system (e.g.~Carroll et al. 1991). 
The spin axis precession of the disc changes two angles: 
1) inclination of the disc to the line of sight (which causes changes 
of the projected area of the disc) and 2) inclination of the projected 
disc to the orbital movement direction (tilt).
As a result we observe varying depth of the minima as it is shown in Fig.~4 
in MG. 
Another kind of a precession -- the nodal regression -- was recently proposed also for
two relatively close eclipsing binaries: SS Lac (Torres 2001) and V907 Sco 
(Lacy et al.~1999) as a explanation of a varying eclipse depth. In both
cases there are precession motion (timescale of
hundred years) of the whole short-period (several days) binary orbit 
caused by an interaction with a third body moving on a wider orbit with a period
of hundreds days. In the case of the long period binary EE Cep, however, we 
observe only the precession of the disc itself. 

\begin{figure}
\begin{minipage}{\linewidth}
\epsfig{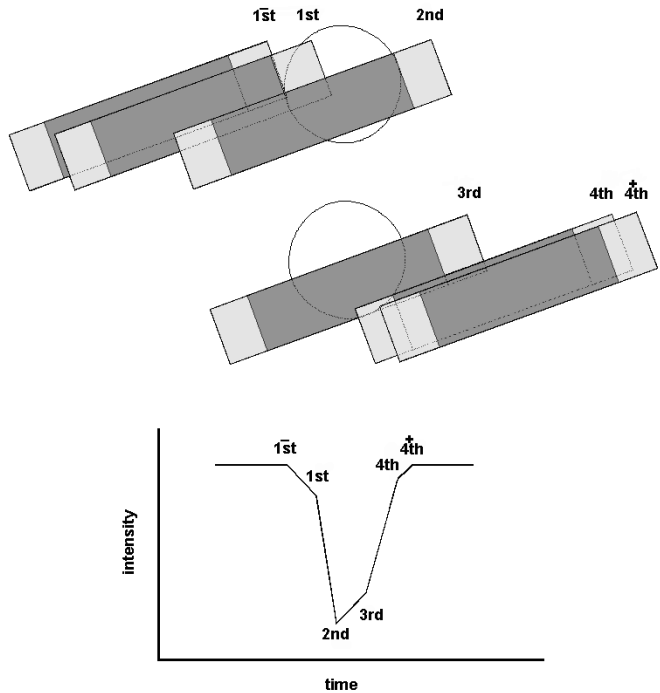}
\caption{The schematic explanation of the characteristic moments observed in the light curve of EE Cep during minima via the disc model. An 'edge-on' position for the disc was assumed for clarity. The B5 primary is marked by a large circle and semitransparent parts of the disc by a light shadow. Because of the disc tilt during the egress the wing may be much shorter or even unobserved -- like during eclipse of epoch 8.}
\label{fig:kota}
\end{minipage}
\end{figure}

During our quantitative analysis and after doing some simple numerical
tests, we have found that the most probable reason for the observed
asymmetries was not the asymmetry of the disc itself but the tilt of the
projected disc to the direction of motion. The tilt produces typical
asymmetry of the minima as the result of the sloping-bottom between inner
contacts (Fig.~\ref{fig:kota}). The wide wings between $1\!^-\!\!$st and
1st, and between 4th and $4\!^+\!$th may be caused by the outer,
semitransparent part of the disc. The ``true'' eclipse is caused by the
opaque, inner part: the four characteristic moments 1st--4th may be easily
understood as contacts between the outer and inner parts of the tilted disc
with the eclipsed star -- see Fig.~\ref{fig:term.ps} and~\ref{fig:kota}.
During the extreme, rare, edge-on and small tilt configuration the eclipses
should have a flat bottom phase as was observed in 1969. The possible period
of nine epochs ($\sim$ 50 years) may reflect precession of the disc.

\section{Spectroscopic behaviour}
 
\begin{figure}
\centering
\includegraphics[width=89mm,height=80mm]{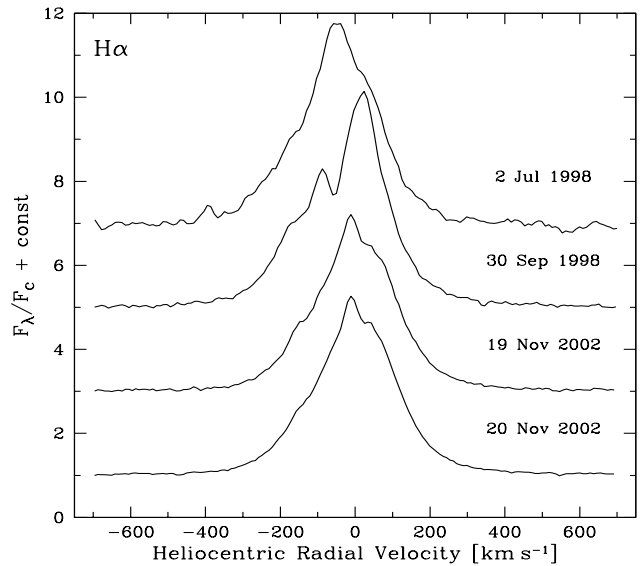}
\caption{$H_\alpha$ profile variations in EE Cep spectra obtained outside eclipse.}
\label{fig:alfa}
\end{figure}

There are very few spectroscopic observations of EE Cep described in the
literature. Herbig (\cite{her60}) gave its spectral type as B5:ne$\beta$.
Br{\"u}ckner (\cite{bryk76}) noted that the emission profile in $H_{\rm \alpha}$
is not typical of Be stars and during the eclipse of epoch 4 there was a
considerable strengthening of this emission in respect to the continuum.
However, his observations were done in very low resolution (200 \AA/mm).
Baldinelli et al.~(\cite{bal81}) did not record any changes in the spectrum
during the following eclipse (E=5), in spite of the slightly higher
resolution (60 \AA/mm). Moreover, they noted broad absorption in the Balmer
series from $H_{\rm \alpha}$ to $H_{8}$ with strong, broad ($H_{\rm \alpha}$) and narrow
($H_{\rm \beta}$) emission lines superimposed. This spectrum was estimated as a
typical Be star of B5 III spectral type, which agrees with the $B-V$, $U-B$
colours analysis (MG). Thus, the connection between Balmer emission and
eclipsed (Be star) or eclipsing (disc) body is not clear.
 
We have secured several spectra covering the $H_{\rm \alpha}$ and the sodium
doublet Na I regions of EE Cep. The spectra were taken at NAO Rozen in
Bulgaria on a 2-m telescope equipped with a coude-spectrograph + LN2-cooled
CCD camera at resolution 0.4 \AA. Unexpected large changes of the $H_{\rm \alpha}$
profile can be noted outside the eclipse phases on a time scale of several
months to years (Fig.~\ref{fig:alfa}). The $H_{\rm \alpha}$ profiles are strong
and broad ($\sim 600$ km/s), show one or two peaked emission with variable
asymmetries. Two profiles obtained during two consecutive days are almost
identical, however a red-shifted bulge around 60 km/s seems to be stronger
in the later spectrum. The sodium doublet D$_1$D$_2$ shows strong
single-component, probably interstellar absorptions. However, its
heliocentric radial velocity probably changed from $V_{\rm r} = -36$ km/s in
September 1998 to $V_{\rm r} = -24$ km/s in November 2002. Also the
equivalent width of D$_1$D$_2$ doublet seems to be variable, decreasing
about 10\% between September 1998 and November 2002. The mean value of the
D$_1$ equivalent width $EW = 0.57$ \AA\ corresponds to $E_{\rm B-V} = 0.46$
using the calibration given by Munari \& Zwitter (\cite{mun97}), which is
fairly consistent with the reddening derived from $UBV$ photometry (MG).

\section{The next eclipse: 2003 ($E=9$)}

The moment of the mid-eclipse according to the elements given in MG is JD
2452794 (3 June 2003). As shown in Fig.~\ref{fig:korel} the eclipse can
start even about 30 days before mid-eclipse. The reason for it is the
presence of the wide wing during ingress which can be usually observed, thus
the photometric observations should start at least 40 days before the
expected mid-eclipse. The brightness of the star should decrease by at least
1 mag in $B,V$ filters, but the observers should keep in mind that this
decrease may be much larger: up to 2 mag! Also, a significantly shallower
(as epochs 3 and 7) minimum can occur. If the periodicity of nine epochs
suggested by Fig.~\ref{fig:order} is real we can expect an eclipse similar
to that of E=0, i.e.~narrow and deep. Near-far infrared photometry during
the eclipses will be very important as it would give the answer to what is
the temperature of the companion.  Also, good quality $B, V$ or $R$ light
curves from the whole eclipse will be valuable to make numerical tests of
the disc model.

The spectroscopic observations appear to be very important. We should expect
the discovery of a shell spectrum produced by the gaseous, semitransparent
part of the disc like in $\epsilon$ Aur (Ferluga \& Mangiacapra 1991).
Moreover, such observations would give us an additional clue to the answer
of the origin of the variable $H_\alpha$ line. The systematic observations
would be very useful to obtain the spectroscopic orbit of the visible B5
component.

The eclipsing body in EE Cep system seems to be unique. The confirmation (or
rejection) of the disc hypothesis can be attained only by detailed
photometric and spectroscopic study of the incoming eclipses.
    
\begin{acknowledgements}
This paper was supported by KBN Grant No.~5 P03D 003 20 and BAN/PAN Joint Research Project No.28. We would like to thank: Ms~Karolina~Wojtkowska for her help in obtaining some rare literature, dr Alojzy~Burnicki for his help in translation of some German publications and dr Boud~Roukema for the very careful reading of the manuscript and for the English corrections in the text. 
\end{acknowledgements}

\bibliographystyle{aa}
%\bibliography{eecep_aa}

\end{document}